\documentclass[journal]{IEEEtran}
\usepackage{color}
\usepackage{amsfonts}
\usepackage{amsmath}
\usepackage{multirow}
\usepackage{graphicx}
\usepackage{amsfonts}
\usepackage{amsmath,epsfig}
\usepackage{stfloats}
\usepackage{amssymb}
\usepackage{url}
\usepackage{bm}
\usepackage{cite}
\usepackage{amsmath}
\usepackage{amssymb}
\usepackage{latexsym}
\usepackage{multirow}
\usepackage{epsfig}
\usepackage{graphics}
\usepackage{mathrsfs}
\usepackage{algorithmic}
\usepackage{algorithm}
\usepackage{subfigure}
\usepackage[all]{xy}

\usepackage{pifont}
\usepackage{bbding}
\usepackage{stmaryrd}
\usepackage{amssymb}
\usepackage{amsfonts}
\usepackage{epic}
\usepackage{stfloats}
\usepackage{latexsym}
\usepackage{epstopdf}
\usepackage{epic}
\usepackage{bm}
\usepackage{xcolor}
\usepackage{mathrsfs}
\usepackage{pifont}
\usepackage{bbding}
\usepackage{amsmath,epsfig}
\usepackage{stmaryrd}
\usepackage{amssymb}
\usepackage{amsfonts}
\usepackage{epic}
\usepackage{graphicx}
\usepackage{subfigure}

\usepackage{enumerate}

\usepackage{stfloats}
\usepackage{latexsym}
\usepackage{epstopdf}
\usepackage{epic}
\usepackage{multirow}
\usepackage{stfloats}
\usepackage{bm}
\usepackage{amsmath}
\usepackage{amssymb}
\usepackage{amsthm}

\usepackage{booktabs}
\usepackage{color}
\usepackage{cases}
\usepackage{array}
\usepackage{makecell}
\usepackage{times}

\usepackage{url}

\usepackage{threeparttable}
\theoremstyle{definition}

\newcommand{\bs}{\boldsymbol}
\usepackage{setspace}
\usepackage{clrscode}
\usepackage{amsmath}
\usepackage{amssymb}
\usepackage{graphicx}
\usepackage{epsfig}
\usepackage{hyperref}
\usepackage{bm}
\usepackage{subfigure}
\usepackage{verbatim}
\usepackage{titletoc}
\usepackage{extarrows}
\usepackage{fancyhdr}
\usepackage{booktabs}
\usepackage{mathrsfs}

\usepackage{lipsum}
\usepackage{cuted}
\usepackage{stfloats}

\usepackage{algorithm}
\usepackage{algorithmic}

\usepackage{balance}
\newcommand{\teq}{\triangleq}
\newcommand{\mbf}{\mathbf}
\newcommand{\mbb}{\mathbb}
\newcommand{\mcal}{\mathcal}
\newcommand{\tbf}{\textbf}

\newcommand{\tr}{\text{Tr}}

\newcommand{\IRS}{\bm{\Theta}}
\newcommand{\bg}{\bs{g}}
\newcommand{\bh}{\bs{h}}
\newcommand{\bq}{\bs{q}}
\newcommand{\bQ}{\bs{Q}}
\newcommand{\bv}{\bs{v}}
\newcommand{\bV}{\bs{V}}
\newcommand{\bH}{\bs{H}}
\newcommand{\EH}{\text{EH}}

\newcommand{\E}{\text{E}}
\newcommand{\I}{\text{I}}
\newcommand{\noi}{\sigma^2}
\newcommand{\pset}{\{ p_k \}}
\newcommand{\fset}{\{ f_k \}}
\newcommand{\df}{\widehat{f}}
\newcommand{\dbf}{\widehat{\bs{f}}}
\newcommand{\dt}{\widehat{\tau}}
\newcommand{\dv}{\widehat{\bv}}
\newcommand{\dV}{\widehat{\bV}}

\newenvironment{sequation}{\begin{equation}\small}{\end{equation}}

\begin{document}
	
	\title{ {Energy Minimization for IRS-aided WPCNs with Non-linear Energy Harvesting Model} }
	%\title{  \textcolor{blue}{...} }
	\author{
		\IEEEauthorblockN{Piao Zeng, Qingqing Wu, \emph{Member, IEEE}, and Deli Qiao} 
		\thanks{P. Zeng and D. Qiao are with the School of Communication and Electronic Engineering, East China Normal University, Shanghai, 200241, China (e-mail:52181214005@stu.ecnu.edu.cn; dlqiao@ce.ecnu.edu.cn). 
		Q. Wu is with the State Key Laboratory of Internet of Things for Smart City, University of Macau, Macau, 999078, China (email: qingqingwu@um.edu.mo). P. Zeng is also with the State Key Laboratory of Internet of Things for Smart City, University of Macau, Macau, China.
		D. Qiao and Q. Wu are also with the National Mobile Communications Research Laboratory, Southeast University, Nanjing, 210096, China.
		%This work is supported in part by the Shanghai Rising-Star Program (21QA1402700), and the FDCT under Grant 0119/2020/A3 and 0108/2020/A, and the Guangdong NSF under Grant 2021A1515011900, and the Open Research Fund of National Mobile Communications Research Laboratory, Southeast University (No. 2020D02 and No. 2021D15). 
	}
	}
	\maketitle
	%
	%\vspace{-10cm}
	% 
	\begin{abstract}
		This paper considers an intelligent reflecting surface(IRS)-aided wireless powered communication network (WPCN), where devices first harvest energy from a power station (PS) in the downlink (DL) and then transmit information using non-orthogonal multiple access (NOMA) to a data sink in the uplink (UL). However, most existing works on WPCNs adopted the simplified linear energy-harvesting model and also cannot guarantee strict user quality-of-service requirements. To address these issues, we aim to minimize the total transmit energy consumption at the PS by jointly optimizing the resource allocation and IRS phase shifts over time, subject to the minimum throughput requirements of all devices. The formulated problem is decomposed into two subproblems, and solved iteratively in an alternative manner by employing difference of convex functions programming, successive convex approximation, and penalty-based algorithm. Numerical results demonstrate the significant performance gains achieved by the proposed algorithm over benchmark schemes and reveal the benefits of integrating IRS into WPCNs. In particular, employing different IRS phase shifts over UL and DL outperforms the case with static IRS beamforming.  
	\end{abstract}
	%\vspace{-1.5mm}
	%\newpage
	
	\vspace{-0.1cm}
	\section{introduction}
	
	The advent of intelligent reflecting surface (IRS) reshapes the wireless propagation environment between transceivers by tuning the reflecting elements, which is able to improve not only the efficiency of wireless energy transmission (WET) but also that of wireless information transmission (WIT) for future Internet-of-Things (IoT) applications \cite{clerckx2021wireless}. This thus has motivated recent research on IRS-aided wireless powered communication networks (WPCNs) and other related applications \cite{lyubin2021IRS,zheng2020joint,wu2021irs,bai2021resource,pan2020intelligent}. 
	In the meanwhile, to further improve the spectral efficiency (SE) as well as user (UE) fairness, non-orthogonal multiple access (NOMA) has been recently employed for uplink (UL) WIT in WPCNs, which allows multiple users to utilize the same spectrum simultaneously by adopting successive interference cancellation (SIC) at the receiver \cite{diamantoulakis2016wireless}. A recent related work proved that dynamic IRS beamforming (adjusting the IRS phase-shift vector over downlink (DL) and UL) was not needed for the  IRS-aided WPCN with a hybrid access point (HAP), which helps reduce the number of IRS phase shifts to be optimized \cite{wu2021irs}. 
	
	\begin{figure}
		\setlength{\abovecaptionskip} {0.cm}
		\setlength{\belowcaptionskip} {-1cm}
		\centering
		\includegraphics[width=0.45\textwidth]{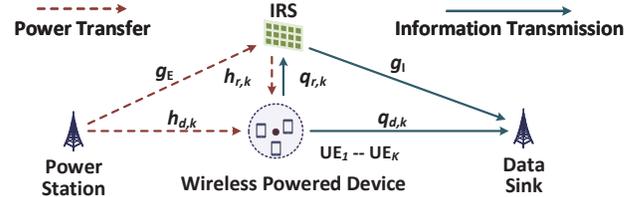}
		\caption{An IRS-aided WPCN.}
		\label{fig:Channel}
	\end{figure}
	
		Nevertheless, most prior works on WPCNs focused on maximizing the sum throughput \cite{lyubin2021IRS,zheng2020joint,wu2021irs}, which neglect the fairness and quality-of-service (QoS) requirements among users. Moreover, they assumed an oversimplified linear energy harvesting (EH) model, which may cause severe resource allocation mismatches and lead to significant performance degradation for practical systems with highly non-linear EH circuits \cite{valenta2014harvesting}. To our best knowledge, the QoS guaranteed resource allocation and beamforming design in the IRS-aided wireless powered NOMA with non-linear EH model have not been investigated yet, which may yield new insights and conclusions different from those in existing works. In particular, it remains unclear whether the dynamic IRS beamforming can improve the system performance or not for power-efficient WPCNs. 
	
	Motivated by the above considerations, we consider an IRS-aided WPCN where an IRS is deployed to assist the DL WET and UL WIT. 
	By employing the separated power station (PS) and data sink (DS) as shown in Fig. \ref{fig:Channel}, the “doubly near-far phenomenon” in WPCNs with a HAP can be avoided \cite{ju2013throughput}. Specifically, our objective is to minimize the system transmit energy consumption at the PS via joint optimization of resource allocation and IRS phase shifts. In particular, minimum throughput requirements are enforced for all the devices and the practical non-linear EH model is incorporated. By decomposing the coupled joint optimization into two subproblems, we solve the energy minimization problem by applying the alternating optimization (AO) technique and propose efficient algorithms for each subproblem respectively. Specifically, we first transform the subproblem of resource allocation into the difference of convex functions (DC) programming and solve it using the successive convex approximation (SCA) technique. On the other hand, for the IRS beamforming optimization, a penalty based algorithm is invoked.
	Numerical results demonstrate the superiority of the proposed algorithms compared to benchmark schemes and reveal the significant benefits of integrating IRS into WPCNs. Moreover, it is found that: 1) More transmit energy consumption can be saved by adopting dynamic IRS reconfigurations over DL and UL compared to the static IRS beamforming, which is different from the result for the throughput maximization problem in \cite{wu2021irs}; 2) Increasing the number of low-cost IRS reflecting elements instead of raising the emitted power at the PS is a more energy-efficient approach for achieving sustainable and green WPCNs. 
	
	\emph{Notations:} Throughout the paper, superscripts $(\cdot)^{T}$, $(\cdot)^{H}$ and \text{diag}$(\cdot)$ represent the transpose,  Hermitian transpose and diagonalization operator, respectively. $\mbb{C}^{a \times b}$ denotes the space of $a \times b$ complex matrices. $|\cdot|$ denotes the absolute value of a complex scalar.  $\|\mbf{X}\|$, $\tr(\mbf{X})$ and \text{Rank}$(\mbf{X})$ denote the Euclidean norm, trace and rank of the matrix $\mbf{X}$, respectively. ${\bf 1}_{N}$ denotes an all-one $N \times 1$ vector.

	\vspace{-0.25cm}
	\section{System Model and Problem Formulation}

	\vspace{-0.08cm}
	\subsection{Channel Model}
	\vspace{-0.08cm}
	As shown in Fig. \ref{fig:Channel}, we consider an IRS-aided WPCN, which consists of one PS, one IRS, one DS, and $K$ IoT devices. We assume that the PS, DS, and all the devices are equipped with a single antenna, while the IRS has $N$ reflecting elements. The total available transmission time is denoted by $T_{\max}$. 
	To characterize the maximum achievable performance, it is assumed that the channel state information (CSI) of all channels is perfectly known at the PS and the DS (see \cite{wu2019towards} and its follow-up works for practical channel acquisition methods). 

	Denote $\bg_\E \in \mbb{C}^{N \times 1}$, $\bh_{r,k} \in \mbb{C}^{N \times 1}$ and $h_{d,k} \in \mbb{C}$, $\forall k \in \mcal{K}\teq\{1,\cdots,K\}$, as the equivalent baseband channels from the PS to the IRS, from the IRS to device $k$, and from the PS to device $k$, respectively. Likewise, the equivalent baseband channels from the IRS to the DS, from device $k$ to the IRS, and from device $k$ to the DS are denoted by $\bg_\I \in \mbb{C}^{N \times 1}$, $\bq_{r,k} \in \mbb{C}^{N \times 1}$ and $q_{d,k} \in \mbb{C}$, $\forall k \in \mcal{K} $, respectively.
	The reflection phase-shift matrix for DL WET and UL WIT are denoted by $\IRS_0 = \text{diag}(e^{j\phi_1},\cdots,e^{j\phi_N})$ and $\IRS_1 = \text{diag}(e^{j\varphi_1},\cdots,e^{j\varphi_N})$, respectively, where $\phi_n$ and $\varphi_n  \in [0,2\pi), \forall n \in \mathcal{N}\teq\{1,\cdots,N\}$. 
	We assume that the WPCN adopts the typical ``harvest and then transmit'' protocol, where each device first collects energy from the radio frequency (RF) signal  broadcasted by the PS with constant transmit power $P_\E$ for a time duration of $\tau_0$ in DL WET. Then, all the devices transmit their information signals to the DS simultaneously for a duration of $\tau_1$ with transmit power $p_k$, $\forall k \in \mcal{K}$, in UL WIT. The RF power received at device $k$ is then given by  
	\begin{sequation}
		\begin{aligned}
			P_{{\EH}_k} &= P_\E | h_{d,k} + \bh_{r,k}^{H} \IRS_0 \bg_\E |^2 = P_\E |h_{d,k} + \bh_k^H \bv_0|^2.
		\end{aligned}
		\label{eq:P_EHk}
	\end{sequation}\noindent
	where $\bh_k^H \teq \bh_{r,k}^{H} \text{diag} (\bg_\E)$, and $\bv_0 \teq [e^{j\phi_1},\cdots,e^{j\phi_N}]^{T}$.
	
	\subsubsection{EH Model}
	To avoid the resource allocation mismatches associated with the traditional linear EH model, we adopt a more general and practical non-linear EH model, which is given by \cite{boshkovska2015practical}
	\begin{sequation}
		\begin{aligned}
			{\Phi}^{\text{non-linear}}_{{\EH}_k} &= \frac{\Psi^{\text{non-linear}}_{{\EH}_k} - M_k \Omega_k}{1-\Omega_k},  \\
			{\Psi}^{\text{non-linear}}_{{\EH}_k} &= \frac{M_k}{1+e^{-a_k(P_{\EH_k}-b_k)}}, \hspace{0.2cm}
			\Omega_k = \frac{1}{1+e^{a_k b_k}},
		\end{aligned}
		\label{eq:nonlinear_P}
	\end{sequation}\noindent
	where ${\Phi}^{\text{non-linear}}_{{\EH}_k}$ denotes the amount of harvested energy at the $k^{th}$ device, and $\Psi^{\text{non-linear}}_{{\EH}_k}$ is the conventional logistic function with respect to the received RF power $P_{{\EH}_k}$. By adjusting the parameters $M_k$, $a_k$ and $b_k$, such a non-linear EH model is able to characterize the joint effects of various non-linear phenomena caused by hardware limitations \cite{boshkovska2015practical}.

	%\vspace{-0.3cm}
	\subsubsection{NOMA Strategy}
	To improve SE and user fairness, NOMA technology is adopted for UL WIT 
	with the help of SIC at the DS. 
	For exposition purpose, we assume that the DS decodes devices' signals based on the indexes from 1 to $K$ in this paper, where the order of indexes are sorted according to the channel gain of direct link, i.e., $q_{d,k}$,  while the proposed algorithm and results are applicable to any given SIC order in practice.  
	Thus, the achievable throughput of the $k$-th device in bits/Hz is given by \cite{wu2021irs} 
	\begin{sequation}
		\begin{aligned}
			r_k = \tau_1 \log_2 \bigg( 1+ \frac{|q_{d,k} + \bq_k^H \bv_1|^2 p_k}{\sum\limits^K_{i=k+1} |q_{d,i} + \bq_i^H \bv_1|^2 p_i + \noi} \bigg) ,
		\end{aligned}
		\label{eq:rk}
	\end{sequation}where $\bq_k^H \teq \bg_\I^{H} \text{diag} (\bq_{r,k})$,  $\bv_1 \teq [e^{j\varphi_1},\cdots,e^{j\varphi_N}]^{T}$, and $\noi$ is the additive white Gaussian noise power at the DS.
	
	\vspace{-0.4cm}
	\subsection{Problem Formulation}
	Different from existing works aiming at maximizing the sum throughput, we pursue a greener transmission policy, which leads to the objective of minimizing the total emitted energy at the PS, i.e., $\tau_0 P_\E$, subject to the total available transmission time and user QoS constraints.
	Note that, since $P_\E$ is usually a given constant due to practical limitation, the objective is equivalent to minimize the time duration $\tau_0$ for the DL WET. Accordingly, the time allocation, the transmit power at the IoT devices and the IRS phase shifts need to be jointly optimized, which leads to the following problem
	
		\vspace{-0.4cm}
		\begin{small}
		\begin{align}
			& { \min_{\scriptstyle \tau_0,\tau_1,\pset, \atop \scriptstyle \bv_0,\bv_1} }  {\tau_0} \label{pro:A}\\
			& \hspace{0.5cm} {\text{s.t.}} \hspace{0.5cm}  { p_k \tau_1 \leq \tau_0 \Phi^{\text{non-linear}}_{{\EH}_k}, \forall k, } \tag{\ref{pro:A}{a}} \label{pro:Aa}\\
			& \hspace{1.35cm} \tau_{0}+\tau_{1} \leq T_{\max } , \tag{\ref{pro:A}{b}} \label{pro:Ab} \\
			& \hspace{1.35cm}  r_k  \geq \Gamma_k, \forall k. \tag{\ref{pro:A}{c}} \label{pro:Af} \\
			& \hspace{1.35cm} \tau_{0} \geq 0, \tau_{1} \geq 0, p_k \geq 0, \forall k , \tag{\ref{pro:A}{d}} \label{pro:Ac} \\
			& \hspace{1.35cm} \left|\left[\bv_0\right]_{n}\right|=1, \forall n \in \mcal{N}, \tag{\ref{pro:A}{e}} \label{pro:Ad} \\
			& \hspace{1.35cm} \left|\left[\bv_1\right]_{n}\right|=1, \forall n \in \mcal{N}. \tag{\ref{pro:A}{f}} \label{pro:Ae} 
		\end{align}
		\end{small}In problem (\ref{pro:A}), (\ref{pro:Aa}) and (\ref{pro:Ab}) represent the energy causality and total time constraints, respectively, (\ref{pro:Af}) are the QoS constraints of IoT devices, (\ref{pro:Ac}) are non-negativity constraints, (\ref{pro:Ad}) and (\ref{pro:Ae}) are unit-modulus constraints of the IRS phase shifts employed for DL WET and UL WIT, respectively. It can be shown that, problem (\ref{pro:A}) is a challenging non-convex optimization problem and difficult to solve optimally in general due to the highly coupled optimization variables in (\ref{pro:Aa}), as well as the non-convex unit-modulus and QoS constraints. In particular, when $\bv_0$ is forced to be the same as $\bv_1$ in problem (\ref{pro:A}), i.e., $\bv_0=\bv_1$, it means that static IRS beamforming is adopted, which will also be investigated for comparison. 
	
	\vspace{-0.3cm}
	\section{Proposed Solution}
	%\vspace{-0.1cm}
	In this section, we first decompose the original problem into two subproblems, namely the resource allocation problem and IRS passive beamforming problem. Then, we propose efficient solutions to the two subproblems respectively. 
	
	\vspace{-0.34cm}
	\subsection{Resource Allocation Optimization}
	\vspace{-0.1cm}
	First, we design the resource allocation strategy for fixed reflection phase-shift vectors $\bv_0$ and $\bv_1$.
	Note that, different from the conclusion in \cite{wu2021irs} where the energy constraint (\ref{pro:Aa}) is met with {\it equality}, letting each device exhaust all its harvested energy may no longer be the optimal solution since each device may not need to use up all the harvested energy to meet with the minimum throughput constraint and a higher transmit power of one device may also cause stronger interference to other devices due to the use of NOMA for UL WIT. Therefore, $\pset$ cannot be eliminated in the optimization as in \cite{wu2021irs}, which are coupled with the time variable $\tau_1$, thus making the problem more difficult to solve. Therefore, we introduce a new set of variables $\fset$, where $f_k = p_k \tau_1, \forall k$. For notation simplicity, define $\widetilde{\bq}_i^H \teq [\bq_i^H \hspace{0.1cm} q_{d,i}]$, $ \widetilde{\bv}_1^H \teq [\bv_1^H \hspace{0.1cm} 1]$, $\bQ_i \teq \widetilde{\bq}_i \widetilde{\bq}_i^H$, $\bV_1 \teq \widetilde{\bv}_1 \widetilde{\bv}_1^H$, and $\bs{f} = [f_1, \cdots, f_K]^T$. The subproblem is reformulated as 
	
	\vspace{-0.4cm}
	\begin{small}
		\begin{align}
			& \min_{\scriptstyle \tau_0,\tau_1,\fset} \tau_0 \label{pro:1.1}\\
			&\hspace{0.45cm} {\text{s.t.}} \hspace{0.2cm}  { f_k  \leq \tau_0 \Phi^{\text{non-linear}}_{{\EH}_k}, \forall k,} \tag{\ref{pro:1.1}{a}} \label{pro:1.1a}\\
			& \hspace{1.cm} (\ref{pro:Ab}), \hspace{0.2cm} \tau_{0} \geq 0, \tau_{1} \geq 0, f_k \geq 0, \forall k , \tag{\ref{pro:1.1}{b}} \label{pro:1.1b} \\
			& \hspace{1.cm}  \tau_1 \log_2 \bigg( 1+ \frac{\tr \big(\bQ_k \bV_1 \big) f_k}{\sum\limits^K_{i=k+1} \tr \big(\bQ_i \bV_1 \big) f_i + \noi \tau_1} \bigg) \notag  \geq \Gamma_k, \forall k. \tag{\ref{pro:1.1}{c}} \label{pro:1.1d} 
		\end{align}
	\end{small}\noindent
	Problem (\ref{pro:1.1}) is still intractable due to the non-convexity of constraint (\ref{pro:1.1d}). To proceed, first we rewrite constraint (\ref{pro:1.1d}) as:
	\begin{sequation}
		\begin{aligned}
			h_k \big(\bs{f}, \tau_1 \big) - g_k \big(\bs{f}, \tau_1 \big) \geq {\Gamma_k}/{\tau_1}, \forall k ,
		\end{aligned}
		\label{eq:c1}
	\end{sequation}\noindent 
	where
	\vspace{-0.4cm}
	\begin{small}
		\begin{align}
			&h_k \big(\bs{f}, \tau_1 \big) \teq \log_2 \bigg( \sum^K_{i=k} \frac{\tr \big(\bQ_i \bV_1 \big)}{\noi} f_i + \tau_1 \bigg), \label{eq:c1.1.1}\\
			&g_k \big(\bs{f}, \tau_1 \big) \teq \log_2 \bigg( \sum^K_{i=k+1} \frac{\tr \big(\bQ_i \bV_1 \big)}{\noi} f_i + \tau_1 \bigg) , \label{eq:c1.2.2}
		\end{align}
	\end{small}\noindent 
	which yield to the classic DC form \cite{zargari2021max}. Note that $g_k \big(\bs{f}, \tau_1 \big)$ is a differentiable concave function with respect to both $\bs{f}$ and $\tau_1$, and is globally upper-bounded by its first-order Taylor expansion at any point. Therefore, we can obtain the upper bound of $g_k \big(\bs{f}, \tau_1 \big)$ at fixed local points $\dbf$ and $\dt_1$ as
	\begin{sequation}
		\begin{aligned}
			g_k \big(\bs{f}, \tau_1 \big) \leq \hspace{0.2cm} & g_k \big(\dbf, \dt_1 \big) + \sum^K_{i=k+1}  \frac{\partial g_k \big(\dbf, \dt_1 \big)}{\partial f_i}  (f_i - \df_i) \\
			& + \frac{\partial g_k \big(\dbf, \dt_1 \big)}{\partial \tau_1}  (\tau_1 - \dt_1) \teq \widehat{g}_k \big(\bs{f}, \tau_1 \big).
		\end{aligned}
	\end{sequation}\noindent 
	By applying the SCA method, problem (\ref{pro:1.1}) can be reformulated in the following convex form
	
	\vspace{-0.5cm}
	\begin{small}
		\begin{align}
			& \min_{\scriptstyle \tau_0,\tau_1,\fset} \tau_0  \label{pro:1.2}\\
			& \hspace{0.5cm} \text{s.t.} \hspace{0.5cm}   \text{(\ref{pro:1.1a}), (\ref{pro:1.1b})}, \hspace{0.2cm}   h_k \big(\bs{f}, \tau_1 \big) - \widehat{g}_k \big(\bs{f}, \tau_1 \big) \geq {\Gamma_k}/{\tau_1}, \forall k , \tag{\ref{pro:1.2}{b}} \label{pro:1.2b}
		\end{align}
	\end{small}\noindent 
	which can be solved by convex optimization solvers such as CVX \cite{GRANT2014CVX}. After obtaining the optimal $\fset$, $\tau_0$ and $\tau_1$, we can readily derive $\pset$ as $p_k = f_k / \tau_1, \forall k$.
	
	\vspace{-0.3cm}
	\subsection{IRS Passive Beamforming Optimization}
	\vspace{-0.1cm}
	Next, we optimize the IRS phase-shift vectors $\bv_0$ and $\bv_1$ for the given $\tau_0$, $\tau_1$, and $\pset$. Note that, since the optimization variables $\bv_0$ and $\bv_1$ are not coupled in the objective and constraints, we can separate this subproblem into two parallel feasibility check problems. 
	
	\subsubsection{DL WET}
	In this stage, our target is to optimize $\bv_0$, subject to (\ref{pro:Aa}) and (\ref{pro:Ad}). Define $\bH_k \teq \widetilde{\bh}_k \widetilde{\bh}_k^H$, $\widetilde{\bh}_k^H \teq [\bh_k^H \hspace{0.2cm} h_{d,k}], \forall k$, $\bV_0 \teq \widetilde{\bv}_0 \widetilde{\bv}_0^H$ and $\widetilde{\bv}_0^H \teq [\bv_0^H \hspace{0.2cm} 1]$. We recast (\ref{pro:Aa}) in a more explicit form which is given by
	\begin{sequation}
		\begin{aligned}
			\tr ( \bH_k \bV_0 ) \geq & \frac{b_k}{P_\E} \! -  \! \frac{1}{a_k P_\E} \ln \! \bigg( \! \frac{M_k}{\frac{1-\Omega_k}{\tau_0}p_k \tau_1+M_k \Omega_k} \! - \! 1 \! \bigg) , \! \forall k.
		\end{aligned}
		\label{eq:c1.4_v0}
	\end{sequation}\noindent 
	To guarantee the rank and unit modulus constraints, the following condition should be satisfied as well.
	\begin{sequation}
		\begin{aligned}
			\text{Rank} ( \bV_0 ) \leq 1, \hspace{0.2cm}
			\bV_0 \succeq 0, \hspace{0.2cm} \text{diag} ( \bV_0 ) = {\bf 1}_{N+1}.
		\end{aligned}
		\label{eq:v0_semi}
	\end{sequation}\noindent 
	Then, the optimal $\bV_0$ can be obtained by solving 
	\begin{small}
		\begin{align}
			&  \text{Find}  \hspace{0.8cm} \bV_0 \label{pro:2.1.3}\\
			& \hspace{0.2cm} {\text{s.t.}} \hspace{0.7cm}  \text{ (\ref{eq:c1.4_v0}), (\ref{eq:v0_semi})} . \tag{\ref{pro:2.1.3}{a}} \label{pro:2.1.3a}
		\end{align}
	\end{small}\noindent 
	Note that most prior works handle such kind of problems by discarding the rank-one constraint and then constructing the approximate solution by applying Gaussian randomization on the high rank solution \cite{wu2019intelligent, wu2019towards, wu2021irs}. However, such a scheme performs far from satisfactory within the AO process. Therefore, we exploit the penalty-based method as in \cite{zargari2021max} instead. Specifically, first we reformulate problem (\ref{pro:2.1.3}) as 
	
	\vspace{-0.5cm}
	\begin{small}
		\begin{align}
			& \ { \min_{\scriptstyle \bV_0} } \hspace{0.8cm} \mu\bigg( \tr ( \bV_0 ) - \| \bV_0 \|_2 \bigg) \label{pro:2.1.4}\\
			& \hspace{0.3cm} {\text{s.t.}} \hspace{0.7cm}   (\ref{eq:c1.4_v0}), \hspace{0.2cm} \bV_0 \succeq 0, \hspace{0.2cm} \text{diag} ( \bV_0 ) = {\bf 1}_{N+1}, \tag{\ref{pro:2.1.4}{a}} \label{pro:2.1.4a}
		\end{align}
	\end{small}\noindent 
	where $\mu$ is the penalty factor. It is worth noting that although the rank-one constraint is relaxed in (\ref{pro:2.1.4}), the solution yields a rank-one solution when $\mu$ is sufficiently large. Nevertheless, (\ref{pro:2.1.4}) is not convex yet due to the DC form of the objective function. To tackle such non-convexity, we utilize the first-order Taylor expansion to obtain the lower bound of $\| \bV_0 \|_2$ and rewrite the objective of (\ref{pro:2.1.4}) as 
	
	\vspace{-0.5cm}
	\begin{small}
		\begin{align}
			 z(\bV_0) \teq \mu\bigg( \tr ( \bV_0 ) - \| \dV_0 \|_2 - \tr \big[ \dv_{\max}\dv_{\max}^H( \bV_0 - \dV_0) \big] \bigg), \label{pro:2.1.5}
		\end{align}
	\end{small}\noindent 
	where $\dv_{\max}$ is the eigenvector corresponding to the maximum eigenvalue of matrix $\dV_0$. Afterwards, we can solve this problem with CVX tool \cite{GRANT2014CVX}.
	
	 To achieve better converged solution, we introduce a ``residual'' variable $\delta$.
	 The rationale is that optimizing the phase shift to enforce the user's equivalent cascaded channel gain to be larger than the original channel gain target in (\ref{pro:2.1.DC.delta.b}) improves the channels' condition, which
	 leads to the reduction of the DL WET time duration $\tau_0$ in problem (\ref{pro:A}). For this consideration, the slack variable $\delta$ can be interpreted as the “channel gain residual” of the users in phase shift optimization, which is beneficial in terms of the convergence manner \cite{wu2019intelligent}. Thus, problem (\ref{pro:2.1.4}) can be further transformed as

	\vspace{-0.4cm}
	\begin{small}
		\begin{align}
			&  \min_{\scriptstyle \bV_0, \delta}  \hspace{0.3cm} z(\bV_0) - \delta  \label{pro:2.1.DC.delta}\\
			& \hspace{0.3cm} {\text{s.t.}} \hspace{0.3cm}  \bV_0 \succeq 0, \hspace{0.2cm} \text{diag} ( \bV_0 ) = {\bf 1}_{N+1},  \tag{\ref{pro:2.1.DC.delta}{a}} \label{pro:2.1.DC.delta.a} \\
			& \hspace{1cm} \tr ( \bH_k \bV_0 ) \geq \delta + \frac{b_k}{P_\E} - \notag \\
			& \hspace{1.6cm} \frac{1}{a_k P_\E} \ln \bigg( \frac{M_k}{\frac{1-\Omega_k}{\tau_0}p_k \tau_1+M_k \Omega_k} -1 \bigg), \forall k. \tag{\ref{pro:2.1.DC.delta}{b}} \label{pro:2.1.DC.delta.b}
		\end{align}
	\end{small}\noindent
	Finally, the optimal $\bV_0$ can be obtained by solving problem (\ref{pro:2.1.DC.delta}) with penalty base method iteratively until the algorithm converges \cite{zargari2021max}. 
	
	\subsubsection{UL WIT}
	To optimize $\bV_1$, first we rewrite the constraint (\ref{pro:Af}) as 
	\vspace{-0.2cm}
	\begin{sequation}
		\begin{aligned}
			& p_k \tr ( \bQ_k \bV_1 )  \geq  \left( \! 2^{\frac{\Gamma_k}{\tau_1}} \! - \! 1 \! \right) \! \bigg(\! \sum\limits^K_{i=k+1} p_i  \tr ( \bQ_i \bV_1 )  + \noi \! \bigg) , \forall k.
		\end{aligned}
		\label{eq:c6.1_v1}
	\end{sequation}\noindent 
	Then, the optimal $\bV_1$ can be obtained similarly as that for problem (\ref{pro:2.1.4}), except replacing the constraint (\ref{eq:c1.4_v0}) by (\ref{eq:c6.1_v1}). Thus, the detail is omitted here for simplicity.
	
	\vspace{-0.3cm}
	\subsection{Static IRS Beamforming}
	To reduce the algorithmic computations and lower the feedback signalling overhead from the PS to the IRS for sending the optimized IRS phase shifts, we consider a special case of employing static IRS beamforming where the IRS phase shifts remain unchanged in DL WET and UL WIT, i.e., $\bV_0=\bV_1\teq \bV$. The optimal IRS passive beamforming can be obtained by solving  
	
	\vspace{-0.4cm}
	\begin{small}
		\begin{align}
			  \text{Find}\hspace{0.1cm}  \hspace{0.2cm} \bV \hspace{6.5cm} & \label{pro:2.2}\\
			  {\text{s.t.}}\hspace{0.2cm}   \tr ( \bH_k \bV ) \geq  \frac{b_k}{P_\E} -  \hspace{4.5cm} & \notag \\ 
			  \hspace{1.8cm}  \frac{1}{a_k P_\E} \ln \! \bigg( \! \frac{M_k}{\frac{1-\Omega_k}{\tau_0}p_k \tau_1+M_k \Omega_k} \! - \! 1 \! \bigg) , \! \forall k, & \tag{\ref{pro:2.2}{a}} \label{pro:2.2a} \\
			 p_k  \tr ( \bQ_k \bV )  \geq  \left( \! 2^{\frac{\Gamma_k}{\tau_1}} \! - \! 1 \! \right) \! \bigg(\! \sum\limits^K_{i=k+1} p_i  \tr ( \bQ_i \bV )  + \noi \! \bigg) , \forall k, & \tag{\ref{pro:2.2}{b}} \label{pro:2.2b} \\
				\text{Rank} ( \bV ) \leq 1, \hspace{0.2cm}
			\bV \succeq 0, \hspace{0.2cm} \text{diag} ( \bV ) = {\bf 1}_{N+1}, \hspace{1.7cm} & \tag{\ref{pro:2.2}{c}}
		\end{align}
	\end{small}\noindent
	which can be tackled by the aforementioned penalty-based method algorithm as well. Note that the number of optimization variables is reduced greatly in this case, especially when the number of IRS elements is practically large, which might be a compromise choice when the computation complexity is a more critical issue in some certain scenarios, striking a balance between the system performance and signaling overhead as well as implementation complexity.
	
	To sum up, the overall joint design problem is solved by alternately optimizing the resource allocation and IRS beamforming until convergence. 
	More rigorously, the convergence of such AO-based algorithm is ensured since the objective value of problem (\ref{pro:A}) is non-increasing over the AO iterations \cite{wu2019intelligent}.
	In addition, the complexity of the AO-based algorithm is analyzed as follows. For the resource allocation subproblem, i.e., problem (\ref{pro:1.1}), the approximate computational complexity is given by $\mathcal{O}(K^{3.5})$. While for the subproblem of IRS passive beamforming optimization, the computational complexity is given by $\mathcal{O}(N^{6.5})$. Thus, the computational complexity of the overall AO algorithm is $\mathcal{O}(I_{AO} (K^{3.5}+N^{6.5}))$, where $I_{AO}$ denotes the number of iterations required for convergence.

	\vspace{-0.2cm}
	\section{Numerical Results}
	
	In this section, we consider an IRS-aided WPCN where the PS, IRS, and DS are located at (-5,0,0) meter (m), (0,0,1) m, and (45,0,0) m, respectively. The IoT devices are randomly and uniformly distributed within a radius of 1 m centered at (0,0,0) m. 
	For large scale fading, the distance-dependent path loss model with respect to the link distance $d$ is given by $PL(d) = c_0(d/d_0)^{-\alpha}$, where $c_0=-30$ dB is the signal attenuation at a reference distance of $d_0 = 1$ m, and $\alpha$ is the pathloss exponent, which is set to be 2.2 for the PS-IRS, IRS-DS and IRS-device channels, while 2.8 for the PS-device and device-DS channels.
	For small scale fading, Rayleigh fading is adopted for PS-device and device-DS direct link, $h_{d,k}$ and $q_{d,k}$, $\forall k$. While for $\bg_\E$, $\bg_\I$, $\bh_{r,k}$ and $\bq_{r,k}$, $\forall k$, Rician fading is adopted with a Rician factor of $10$. The parameters for non-linear EH model are set as $a_k = 150, b_k = 0.014$, and $M_k = 0.024$ \cite{zargari2021max}. The other system parameters are set as follows: $P_{\E}=40$ dBm, $\noi=-
	90$ dBm, $T_{\max}=1$ s, $\Gamma =0.1$ bits/Hz, $K=3$ and $N=100$ \cite{wu2021irs}, if not specified otherwise.
	
	For comparison, we consider the following schemes.  1) \textbf{Proposed AO with dynamic IRS:} employ different IRS phase shifts in DL and UL with the proposed solutions; 
	2) \textbf{GR with dynamic IRS:} where Gaussian randomization is applied to recover the rank-one $\bV_0$ and $\bV_1$ based on the SDR solution \cite{wu2019intelligent}; 3) \textbf{ETA with dynamic IRS:} i.e., $\tau_0=\tau_1$ based on the solution of the scheme in 1); 
	4) \textbf{Proposed AO with static IRS:} use the same IRS phase shifts in DL and UL, i.e., $\bV_0 = \bV_1$ with the proposed solutions; 5) \textbf{Fixed IRS phase shift:} Only optimize resources allocation, i.e., $\tau_0,\tau_1,\pset$, with randomly initialized IRS phase shifts; 6) \textbf{Without IRS.}
	
	\begin{figure*}[!t]
		\centering
		\subfigure[Optimized $\tau_0$ versus $N$.]{
			\label{fig:N_t0}
			\includegraphics[width=0.4885\columnwidth,height=0.487\columnwidth]{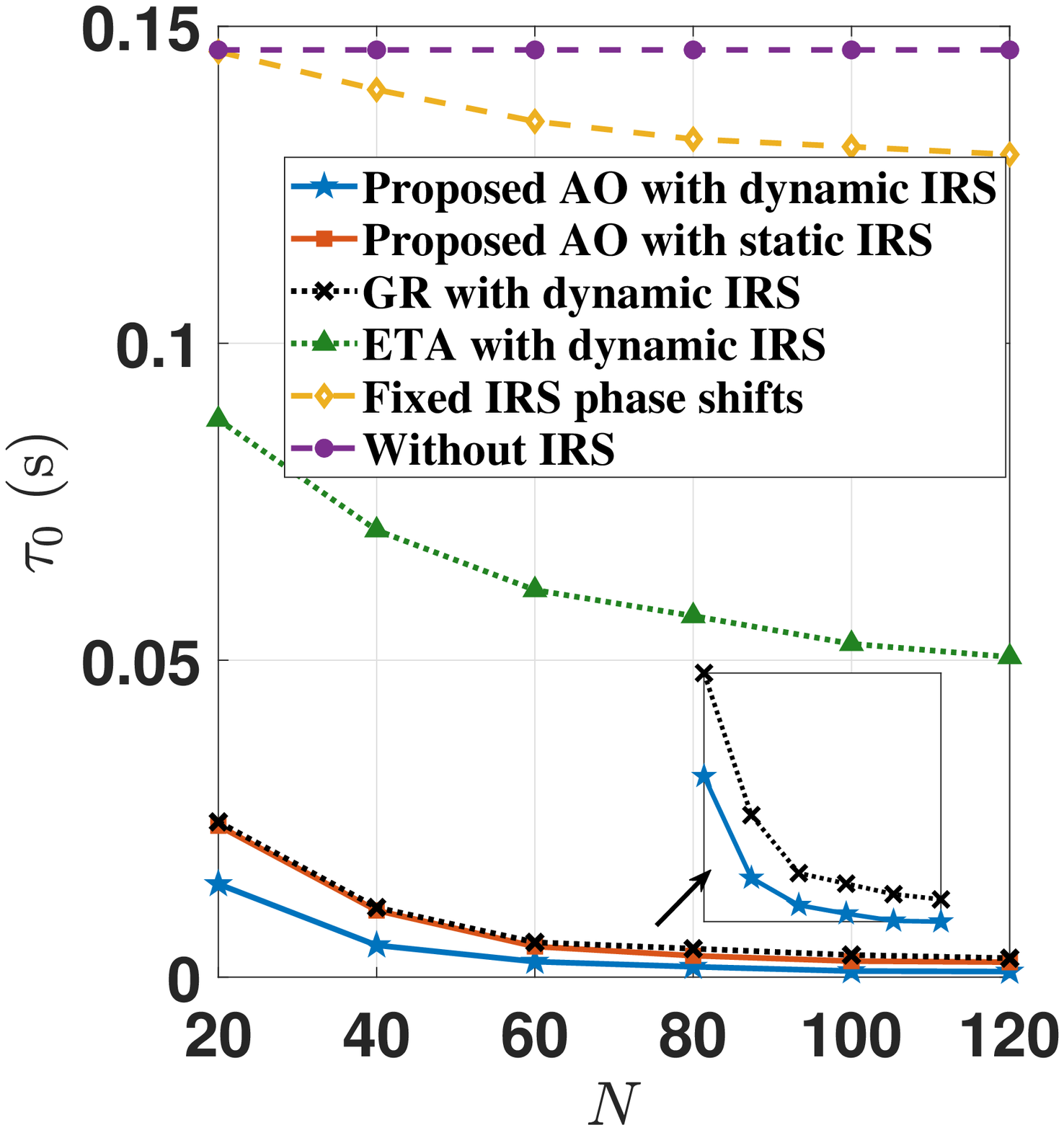}}
		\subfigure[Optimized $\tau_0$ versus $P_\E$.]{
			\label{fig:PA_t0}
			\includegraphics[width=0.4885\columnwidth,height=0.4885\columnwidth]{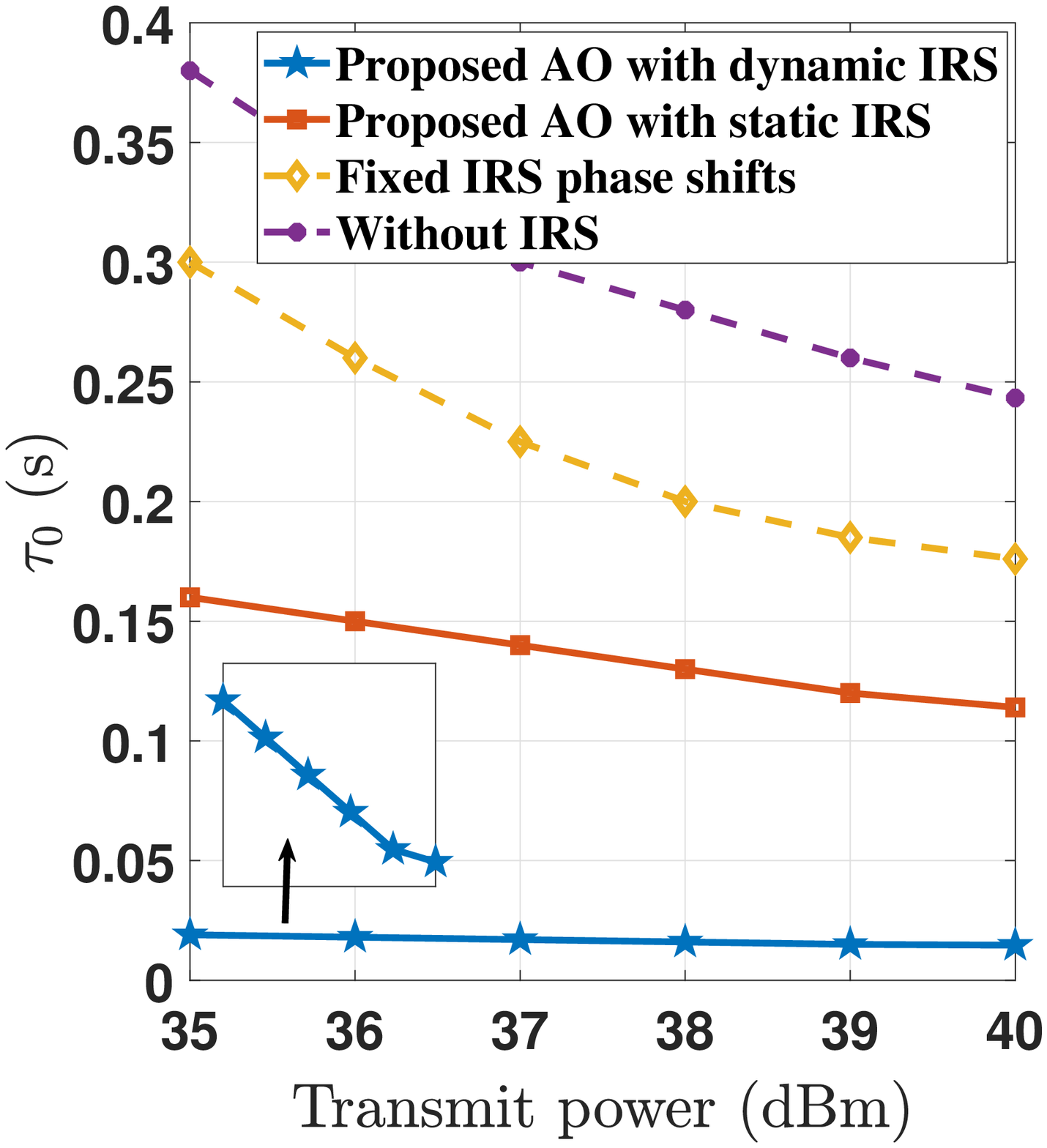}}
		\subfigure[Transmit energy  $\tau_0 P_\E$ versus $P_\E$.]{
			\label{fig:PA_E}
			\includegraphics[width=0.4885\columnwidth,height=0.4885\columnwidth]{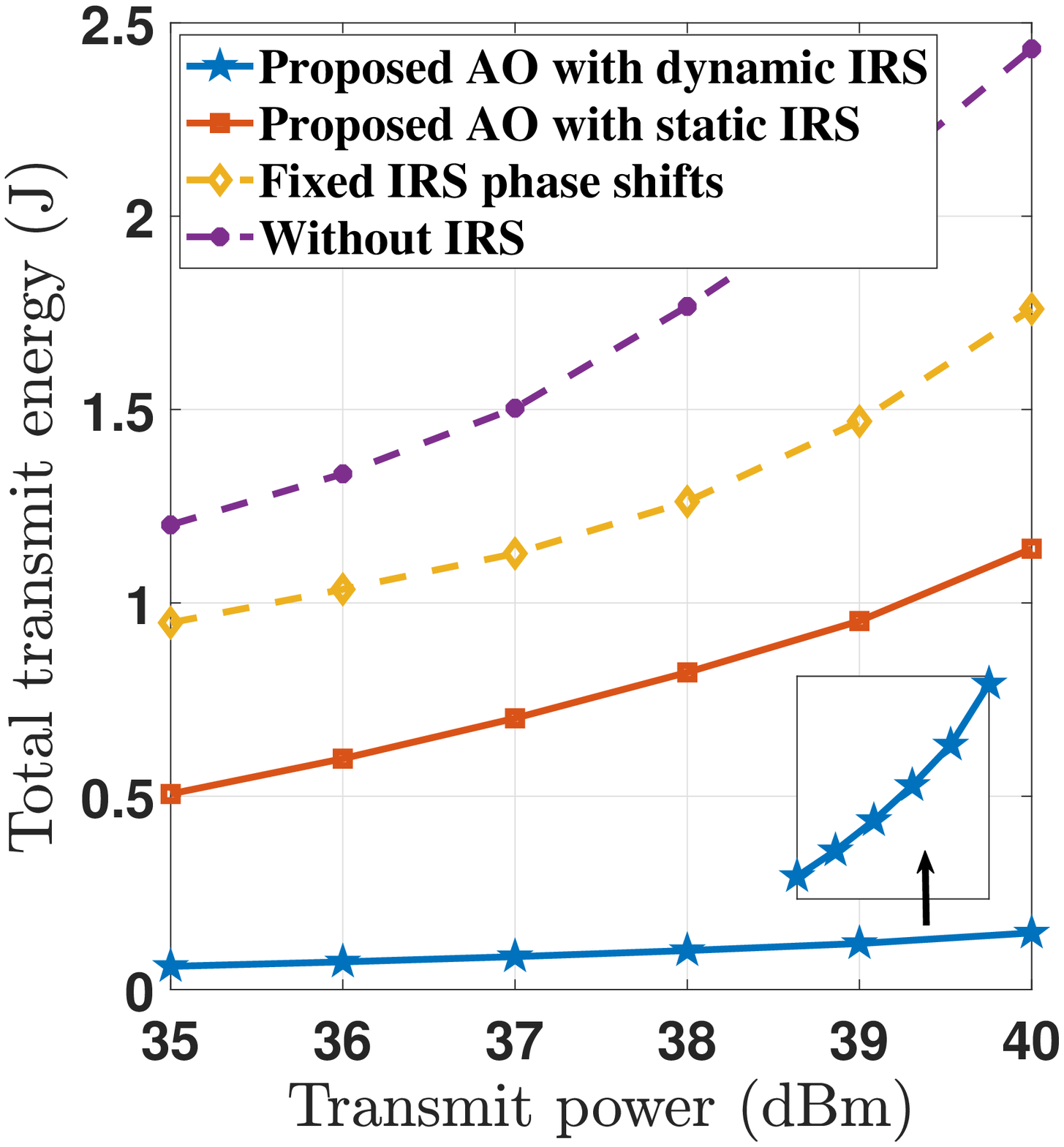}}
		\subfigure[Effect of imperfect CSI.]{
			\label{fig:imCSI}
			\includegraphics[width=0.4885\columnwidth,height=0.5\columnwidth]{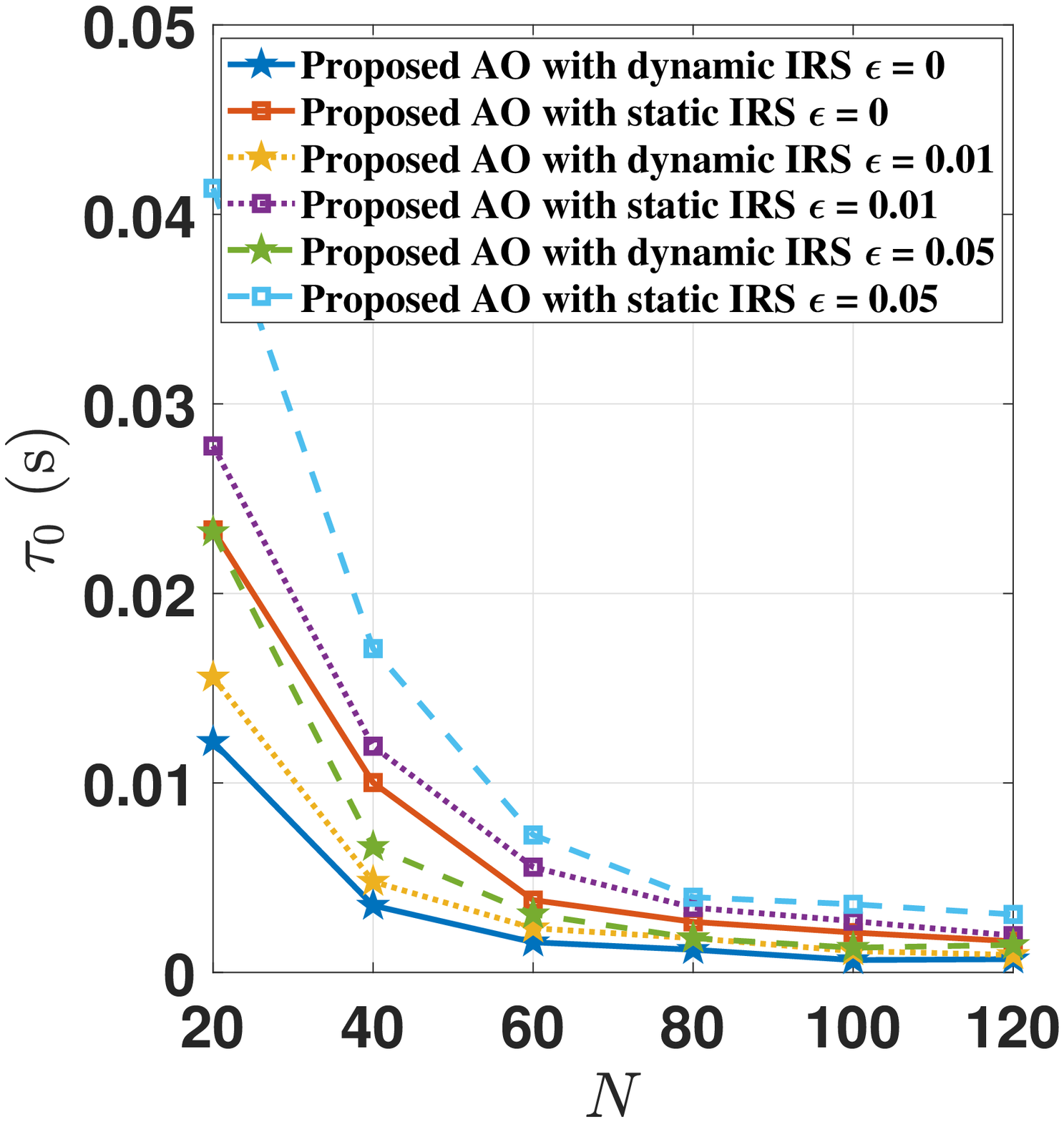}}
			\caption{Simulation results.}
	\end{figure*} %\vspace{-0.1cm}

	First, we investigate the impact of the number of IRS elements. As shown in Fig. \ref{fig:N_t0}, the optimized $\tau_0$ keeps decreasing as the number of IRS elements increases for all the strategies, except the benchmark of without IRS. This is expected since for the IRS-aided WPCN, the increasing number of IRS elements brings more cascade channels, which enables the higher channel gain. Thus, the DL WET duration $\tau_0$ can be lowered, which shows the benefits of IRS for WPCNs since we can reduce the PS's consumed transmit power for DL WET by deploying IRS with a large number of low-cost reflection elements. 
	Besides, it can be seen that when employing dynamic IRS, the optimized $\tau_0$ obtained by the proposed algorithm is much lower compared with that of the system with "GR", "ETA" and "Fixed IRS phase shifts", which reveals the necessity of properly designed joint resource allocation and phase shift optimization. 
	Furthermore, it is observed that the "Proposed AO with dynamic IRS" outperforms "Proposed AO with static IRS", indicating the benefits of taking different IRS configurations over the DL WET and UL WIT, which is different from the conclusion in \cite{wu2021irs}, where using same IRS phase shifts suffices to achieve the optimal performance with a HAP.
	This results from the considered system model and problem formulation.
	In detail, employing separated PS and DS architecture leads to different UL/DL channels. Besides, using the non-linear energy harvesting model in DL WPT further makes DL channels asymmetric as the UL channels.
	This result highlights the advantage of dynamic IRS beamforming as well as the superiority of the proposed design among the benchmarks.

	Fig. \ref{fig:PA_t0} and Fig. \ref{fig:PA_E} present respectively the optimized DL WET duration and the corresponding transmit energy consumption as the transmit power $P_\E$ at the PS varies. Specifically, it is observed from Fig. \ref{fig:PA_t0} that the optimized $\tau_0$ for DL WET declines as the transmit power increases, because increasing the transmit power is able to shorten the energy charging time to meet with the QoS constraint in UL WIT. Nevertheless, we find from Fig. \ref{fig:PA_E} that increasing the transmit power leads to an increase in total transmit energy consumption $\tau_0 P_\E$, indicating that increasing the transmit power is not a favorable choice to reduce the overall transmit energy consumption in practical WPCNs. 
	This is due to the fact that we adopt non-linear EH model in the paper, the harvest energy is not linearly proportional to the increase of the transmit power, which leads to that the reduction of the optimized DL WET time duration $\tau_0$ is not linearly proportional to the increase of the transmit power. From Fig. \ref{fig:PA_E}, it is easy to infer that the increasing of the transmit power is more prominent than the reduction of time $\tau_0$. Therefore, the total transmit energy is increasing with large transmit power.
	By comparing Fig. \ref{fig:N_t0} and Fig. \ref{fig:PA_E}, it is not difficult to conclude that we can save the transmit energy consumption by deploying more low-cost reflecting elements instead of raising the emitted power at the PS, which further demonstrates the benefits of integrating properly designed IRS into WPCNs.

	In addition, we also examine the effect of imperfect CSI on the system performance as shown in Fig.\ref{fig:imCSI}, where $\epsilon$ denotes the value of channel estimation error. 
	It can be seen that, when the available CSIs are imperfect, the performance of the proposed algorithm decreases as expected. And this degradation gets worsen as the channel estimation errors $\epsilon$ are increasing. However, the observations regarding to the comparison between dynamic IRS beamforming and static IRS beamforming in this letter still hold even with estimation errors of CSIs. More importantly, we also observe that as $N$ increases, the performance gap between the perfect CSI and imperfect CSI cases becomes smaller, which implies that the performance loss caused by imperfect CSI can be effectively alleviated by using a large number of reflecting elements and thus validates the practically of the proposed method.

	\vspace{-0.2cm}
	\section{Conclusions}
	%\vspace{-0.3cm}
	In this paper, we investigated the joint passive beamforming and resource allocation optimization for an IRS-aided WPCN employing NOMA for UL WIT. In particular, the total emitted energy at the PS was minimized subject to the total available transmission time and devices' QoS constraints under a non-linear EH model. The joint optimization problem was decomposed into two subproblems and solved respectively in an alternative manner by using the SCA technique and penalty based algorithm. Numerical results not only validated the effectiveness of the proposed AO algorithm but also unveiled the following useful insights. 1) Using dynamic IRS beamforming in DL and UL outperforms employing static IRS beamforming for separated PS and DS deployment. 2) The transmit energy consumption can be effectively reduced by deploying more low-cost IRS reflecting elements instead of raising the emitted power at the PS.

	\vspace{0.05cm}
	
	\bibliographystyle{IEEEtran}
	\bibliography{IEEEabrv,piaobib}

\end{document}